\documentclass[fleqn,usenatbib]{mnras}

\usepackage{newtxtext,newtxmath}
\usepackage[T1]{fontenc}
\usepackage[switch]{lineno}

\DeclareRobustCommand{\VAN}[3]{#2}
\let\VANthebibliography\thebibliography
\def\thebibliography{\DeclareRobustCommand{\VAN}[3]{##3}\VANthebibliography}

\usepackage{adjustbox}
\usepackage{graphicx}	
\usepackage{amsmath}	
\usepackage{float}

\title[Two Dynamically Discovered Compact Objects from LAMOST Survey]{Two Dynamically Discovered Compact Object Candidate Binary Systems from LAMOST Low-resolution Survey}

\author[S. Qi et al.]{
Senyu Qi,$^{1}$
Wei-Min Gu,$^{1}$\thanks{E-mail: guwm@xmu.edu.cn}
Zhi-Xiang Zhang,$^{1}$
Tuan Yi,$^{2}$
Jin-Zhong Liu,$^{3}$
Ling-Lin Zheng$^{1}$
\\
$^{1}$Department of Astronomy, Xiamen University, Xiamen, Fujian 361005, People's Republic of China\\
$^{2}$Department of Astronomy, School of Physics, Peking University, Beijing 100871, People's Republic of China\\
$^{3}$Xinjiang Astronomical Observatory, Chinese Academy of Sciences, Urumqi, Xinjiang 830011, People’s Republic of China
}

\date{Accepted XXX. Received YYY; in original form ZZZ}

\pubyear{2024}

\begin{document}
\label{firstpage}
\pagerange{\pageref{firstpage}--\pageref{lastpage}}
\maketitle

\begin{abstract}
We report two binary systems, LAMOST J035540+381550 (hereafter J035540) and LAMOST J035916+400732 (hereafter J035916), 
identified through the Large Sky Area Multi-Object Fiber Spectroscopic Telescope (LAMOST) low-resolution survey (LRS). 
Each of these two systems contains an M-type star orbiting with a invisible compact object candidate. 
Follow-up spectroscopic observations of Palomar 200-inch telescope (P200) enhance radial velocity measurements. 
We use radial velocities from LAMOST and P200, as well as light curves from Zwicky Transient Facility (ZTF) to constrain orbital parameters. 
The masses of the visible M-type stars are estimated by fitting the MIST isochrones and SEDs.
The mass functions for the unseen companions are: $0.22\pm0.01 M_{\odot}$ for J035540 and $0.16\pm0.01 M_{\odot}$ for J035916. 
With the orbital and stellar parameters derived above and assuming the orbital inclination is 90 degree (edge-on), 
we find that the minimum masses of the invisible companions exceeds that of the visible stars.
The single-lined feature and the dynamical evidence suggest the presence of compact objects.
J035540's ZTF light curve, modeled with \texttt{PHOEBE}, yields a compact object mass of $0.70^{+0.12}_{-0.05} M_{\odot}$. 
For J035916, ellipsoidal modulation analysis constrains the light curve amplitude, 
yielding a compact object mass range of $0.57-0.90 M_{\odot}$.
The mass estimates indicate that both are likely white dwarfs. 
These findings underscore the efficiency of optical time-domain surveys and dynamical methods in identifying faint, massive white dwarfs, along with other compact objects in binaries.
\end{abstract}


\section{Introduction}
\label{sec:intro}

White dwarfs (WDs), neutron stars (NSs), and black holes (BHs) are the products of stellar evolution in the final stages of the life cycle. Numerous observations indicate the presence of a substantial number of compact objects in binary systems \citep{1984ARA&A..22..537J,2006ARA&A..44...49R,2012MNRAS.419..806R}. The searching and identification of these compact objects within binary systems play a crucial role in advancing our understanding of stellar formation, evolutionary history, binary interactions and evolution, as well as the properties and behavior of matter under extreme densities. Concurrently, compact object binary systems are intricately linked to a plethora of high-energy astrophysical phenomena, including novae, supernovae, X-ray bursts, and gravitational waves \citep{1971ARA&A...9..183P,1973ApJ...186.1007W, 1976IAUS...73...75P,1984ApJ...277..355W,1984ApJ...286..644N,2016PhRvL.116f1102A,2017ApJ...848L..12A}.

In recent years, with the rapid development of optical time-domain surveys, the dynamical searching and identification of binary systems containing compact objects have witnessed a significant upsurge. The dynamical method relies on multi-epoch spectral observations to identify stars exhibiting significant radial velocity variations caused by compact companion stars \citep{1969ApJ...156.1013T}. Recently, several stellar-mass BHs and NSs in binary systems have been uncovered utilizing the dynamical method based on astrometric, spectroscopic, and photometric time-domain surveys \citep{2019Sci...366..637T,2019Natur.575..618L,2020ApJ...900...42L,2021MNRAS.504.2577J,2022MNRAS.tmp.2933E,2022NatAs.tmp..201Y,2023ApJ...946...79T,2023MNRAS.521.4323E,2023SCPMA..6629512Z}. In the pursuit of non-interacting binary BH systems, numerous false positives have been confirmed to be actual luminous binary systems \citep{2021MNRAS.502.3436E,2022MNRAS.512.5620E,2022MNRAS.515.1266E}. Simultaneously, the nature of certain NS candidates identified through dynamical methods remains uncertain. This uncertainty arises from measured masses that do not surpass the Chandrasekhar limit, leaving ambiguity regarding whether they are NSs or massive WDs \citep{2023ApJ...944L...4L,2024ApJ...961L..48Z}.

The dynamical method offers a robust approach to search for binary systems with WDs. \citet[][Table 6]{2024MNRAS.529..587R} presents a collection of WD candidates identified using this method. Their table lists 12 WD candidates, with visible companions that include G-, K-, and M-type stars.
Over the past few decades, the sample of white dwarf-main sequence (WD-MS) binaries has significantly increased. The majority of these binaries are identified through the prominent hydrogen and helium absorption lines characteristic of WDs in their blue-end spectra, which are more easily detectable when the WDs are hot \citep{2012MNRAS.419..806R,2014MNRAS.445.1331L,2017ApJ...850...34B,2018MNRAS.477.4641R}. Due to the intrinsic dimness of WDs, search methods relying on WD's unique spectral components are generally confined to nearby sources. Furthermore, a substantial population of young ($\lesssim 1\  \rm{Gyr}$) and hot ($\gtrsim 10000\  \rm{K}$) WDs has been discovered leveraging UV excess \citep{2016MNRAS.463.2125P,2021MNRAS.501.1677H}. However, the galactic population of massive WDs \citep[$\gtrsim 0.8 M_{\odot}$;][]{2022MNRAS.517.2867H} requires additional supplementation. Measuring the mass distribution of WDs holds crucial implications for understanding pulsating phases on the asymptotic giant branch \citep{2009ApJ...693..355W,2018ApJ...866...21C}, stellar population synthesis \citep{1998MNRAS.300..872M}, and galaxy evolution theories \citep{2008MNRAS.387.1693C,2014ApJ...782...17K}. Massive WDs have a more compact structure, resulting in lower luminosity, which makes them challenging to detect. Therefore, dynamical methods present an opportunity to expand the sample of massive WDs.

The Large Sky Area Multi-Object Fiber Spectroscopic Telescope (LAMOST) is a four-meter-aperture, reflective Schmidt telescope with a wide field of view ($\sim 5^\circ$) \citep{2012RAA....12.1197C}. With 4000 fibers, LAMOST is capable of providing millions of stellar spectra in both medium-resolution mode ($R \sim 7500$) and low-resolution mode ($R \sim 1800$). 
As of data release 10 (DR10), LAMOST has released spectra for 22.29 million stars.
Notably, LAMOST's observation strategy involves multiple-epoch observations of the same target, contributing significantly to the application of dynamical methods in the search for binary systems containing compact objects \citep{2019ApJ...872L..20G,2019AJ....158..179Z,2021ApJ...923..226Y,2022SCPMA..6529711M,2022ApJ...938...78L,2022MNRAS.517.4005M,2022ApJ...940..165Y,2023AJ....165..187Q}.

In this work, we present two close binary systems discovered through radial velocity observations, each composed of a white dwarf candidate and an M-type companion. Section \ref{sec:data} describes the discovery of these systems and follow-up spectroscopic observations. In Section \ref{sec:result}, we discuss the properties of the two sources, including the measurement of visible star's parameters and the determination of their orbital parameters. The discussion and summary are provided in Sections \ref{sec:discussion} and \ref{sec: summmary}, respectively.

\section{Observation and data analysis} 
\label{sec:data}

LAMOST J035540+381550 (hereafter J035540) and LAMOST J035916+400732 (hereafter J035916) have been identified as single-lined spectroscopic binaries each consisting of a K/M type companion and an unseen compact object candidate by \citet{2022SCPMA..6529711M} based on the LAMOST low-resolution survey (LRS). J035540 and J035916 were selected due to their significant radial velocity variations. Since the photometric data from TESS of J035540 and J035916 do not show significant photometric periodicity, \citet{2022SCPMA..6529711M} estimated a theoretical lower limit of the orbital period based on the Roche-lobe radius assumption and the mass-radius relation for low-mass main sequence star.

\begin{table*}
\caption{Statistics of the observed spectra of J035540 and J035916. \raggedright \textit{Notes:} \textbf{ID}: the target id;
\textbf{Telescope}: the telescopes used for the observations;
\textbf{HMJD}: the modified heliocentric Julian date at the middle of an exposure;
\textbf{Obs.Data}: the UTC time for the middle of an exposure;
\textbf{Exptime}: duration of each observation;
\textbf{RV(blue)}: the radial velocity measured from the blue side of the spectrum;
\textbf{SNR(blue)}: the signal-to-noise ratio of the blue side of the spectrum;
\textbf{RV(red)}: the radial velocity measured from the red side of the spectrum;
\textbf{SNR(red)}: the signal-to-noise ratio of the red side of the spectrum.
The spectra from Xinglong 2.16m, due to poor quality, did not yield measurable radial velocities.
$^{*}$ The difference between RV(red) and RV(blue) is attributed to atmospheric refraction causing greater displacement in the blue arm of the spectrum at lower zenith angles. During the exposure, the airmass for J035916 was 1.722, corresponding to a zenith angle of $35.5^\circ$. According to \citet[][Figure 8]{2015AJ....150...19L}, we estimate a deviation of $\sim\mathrm{30\  km \ s^{-1}}$ in the blue arm RV measurement due to atmospheric refraction, which is consistent with the deviation between blue and red arms.
}
\label{table:1}
\begin{adjustbox}{width=1\textwidth}
\begin{tabular}{llrlrrrrr}
\hline\hline
ID & Telescope & HMJD & Obs.Date & Exptime & RV(blue) & SNR(blue) & RV(red) & SNR(red)\\
{} & {} & {} & {} & (s)  & ($\mathrm{km\,s^{-1}}$) & {} & ($\mathrm{km\,s^{-1}}$) & {}\\
\hline
J035540 & LAMOST LRS & 57386.535782 & 2015-12-30 12:45:00 & 1800 & 19.1 $\pm$ 6.9 & 8.6 & 7.1 $\pm$ 2.5 & 45.7  \\
J035540 & LAMOST LRS & 57386.559392 & 2015-12-30 13:19:00 & 1800 & 61.5 $\pm$ 7.2 & 7.6 & 59.3 $\pm$ 3.6 & 38.5  \\
J035540 & LAMOST LRS & 57386.582307 & 2015-12-30 13:52:00 & 1800 & 103.9 $\pm$ 6.6 & 8.6 & 95.3 $\pm$ 5.3 & 45.6  \\
J035540 & LAMOST LRS & 57724.615126 & 2016-12-02 14:38:00 & 1800 & -53.3 $\pm$ 6.9 & 11.4 & -49.3 $\pm$ 3.3 & 66.9  \\
J035540 & LAMOST LRS & 57724.638043 & 2016-12-02 15:11:00 & 1800 & -5.2 $\pm$ 6.1 & 9.1 & -4.8 $\pm$ 1.7 & 55.1  \\
J035540 & LAMOST LRS & 57724.660959 & 2016-12-02 15:44:00 & 1800 & 41.6 $\pm$ 5.4 & 11.0 & 34.4 $\pm$ 3.7 & 57.4  \\
J035540 & P200 & 60198.374980 & 2023-09-11 08:58:11 & 1200 & 130.5 $\pm$ 4.2 & 12.9 & 129.1 $\pm$ 1.9 & 28.9  \\
J035540 & P200 & 60198.394428 & 2023-09-11 09:26:11 & 1200 & 105.5 $\pm$ 5.0 & 13.9 & 103.3 $\pm$ 1.8 & 29.1  \\
J035540 & P200 & 60198.460415 & 2023-09-11 11:01:12 & 1200 & -26.6 $\pm$ 5.0 & 14.2 & -32.1 $\pm$ 1.8 & 30.8  \\
J035540 & Xinglong 2.16 m & 60228.662909 & 2023-10-11 15:36:44 & 1500 & ... & 2.8 & ... & 2.8 \\
J035540 & Xinglong 2.16 m & 60228.680376 & 2023-10-11 16:01:53 & 1500 & ... & 3.2 & ... & 3.2 \\ 
J035916 & LAMOST LRS & 57386.535796 & 2015-12-30 12:45:00 & 1800 & -162.7 $\pm$ 9.0 & 8.7 & -160.8 $\pm$ 4.6 & 39.9  \\
J035916 & LAMOST LRS & 57386.559406 & 2015-12-30 13:19:00 & 1800 & -171.8 $\pm$ 7.2 & 7.1 & -174.0 $\pm$ 4.7 & 40.2  \\
J035916 & LAMOST LRS & 57386.582321 & 2015-12-30 13:52:00 & 1800 & -155.2 $\pm$ 10.7 & 7.6 & -155.5 $\pm$ 4.6 & 39.8  \\
J035916 & LAMOST LRS & 57724.615084 & 2016-12-02 14:38:00 & 1800 & -38.0 $\pm$ 8.1 & 9.2 & -14.6 $\pm$ 3.2 & 59.0  \\
J035916 & LAMOST LRS & 57724.638000 & 2016-12-02 15:11:00 & 1800 & -77.3 $\pm$ 12.8 & 7.6 & -63.9 $\pm$ 5.4 & 45.3  \\
J035916 & LAMOST LRS & 57724.660917 & 2016-12-02 15:44:00 & 1800 & -137.3 $\pm$ 8.1 & 8.9 & -127.1 $\pm$ 3.3 & 54.8  \\
J035916 & LAMOST LRS & 59554.619904 & 2021-12-06 14:45:00 & 1800 & -2.8 $\pm$ 13.4 & 6.6 & 6.5 $\pm$ 4.7 & 34.9  \\
J035916 & LAMOST LRS & 59554.642126 & 2021-12-06 15:17:00 & 1800 & 50.3 $\pm$ 14.6 & 6.0 & 62.5 $\pm$ 5.3 & 33.2  \\
J035916 & LAMOST LRS & 59554.664348 & 2021-12-06 15:49:00 & 1800 & 82.0 $\pm$ 10.2 & 6.8 & 105.3 $\pm$ 5.3 & 34.4  \\
J035916 & P200 & 60198.334762 & 2023-09-11 08:00:27 & 1200 & {$^{*}$} -70.1 $\pm$ 6.7 & 10.2 & -45.2 $\pm$ 2.6 & 23.3  \\
J035916 & P200 & 60198.352472 & 2023-09-11 08:25:57 & 1500 & -108.6 $\pm$ 7.0 & 11.2 & -92.8 $\pm$ 2.4 & 27.8  \\
J035916 & P200 & 60198.416755 & 2023-09-11 09:58:30 & 1500 & -176.0 $\pm$ 4.5 & 15.4 & -173.6 $\pm$ 1.9 & 32.5  \\
J035916 & P200 & 60198.439392 & 2023-09-11 10:31:06 & 1500 & -158.7 $\pm$ 3.6 & 17.5 & -157.2 $\pm$ 1.6 & 37.3  \\
J035916 & P200 & 60198.483450 & 2023-09-11 11:34:32 & 1500 & -87.9 $\pm$ 4.4 & 15.6 & -83.7 $\pm$ 1.9 & 33.6  \\
J035916 & P200 & 60198.501530 & 2023-09-11 12:00:34 & 1500 & -41.3 $\pm$ 4.6 & 16.8 & -42.6 $\pm$ 2.1 & 33.4  \\
\hline
\end{tabular}
\end{adjustbox}
\end{table*}


\subsection{LAMOST spectra} 

The LAMOST LRS covers a wavelength range of 3690 to 9100 \r{A} with a spectral resolution of approximately 1800. An LAMOST LRS observation typically involves three consecutive exposures, each lasting 10-30 minutes. Subsequently, the spectra from these exposures are merged to improve the signal-to-noise ratio (SNR). 
The single-exposure spectra are archived in the first data release of LAMOST LRS single-epoch spectra \citep{2021RAA....21..249B}. These spectra consist of two arms: the blue arm covers 3690-6000 \r{A}, and the red arm covers 5700 to 9100 \r{A}. 

\begin{figure*}
\centering
\includegraphics[scale=0.28]{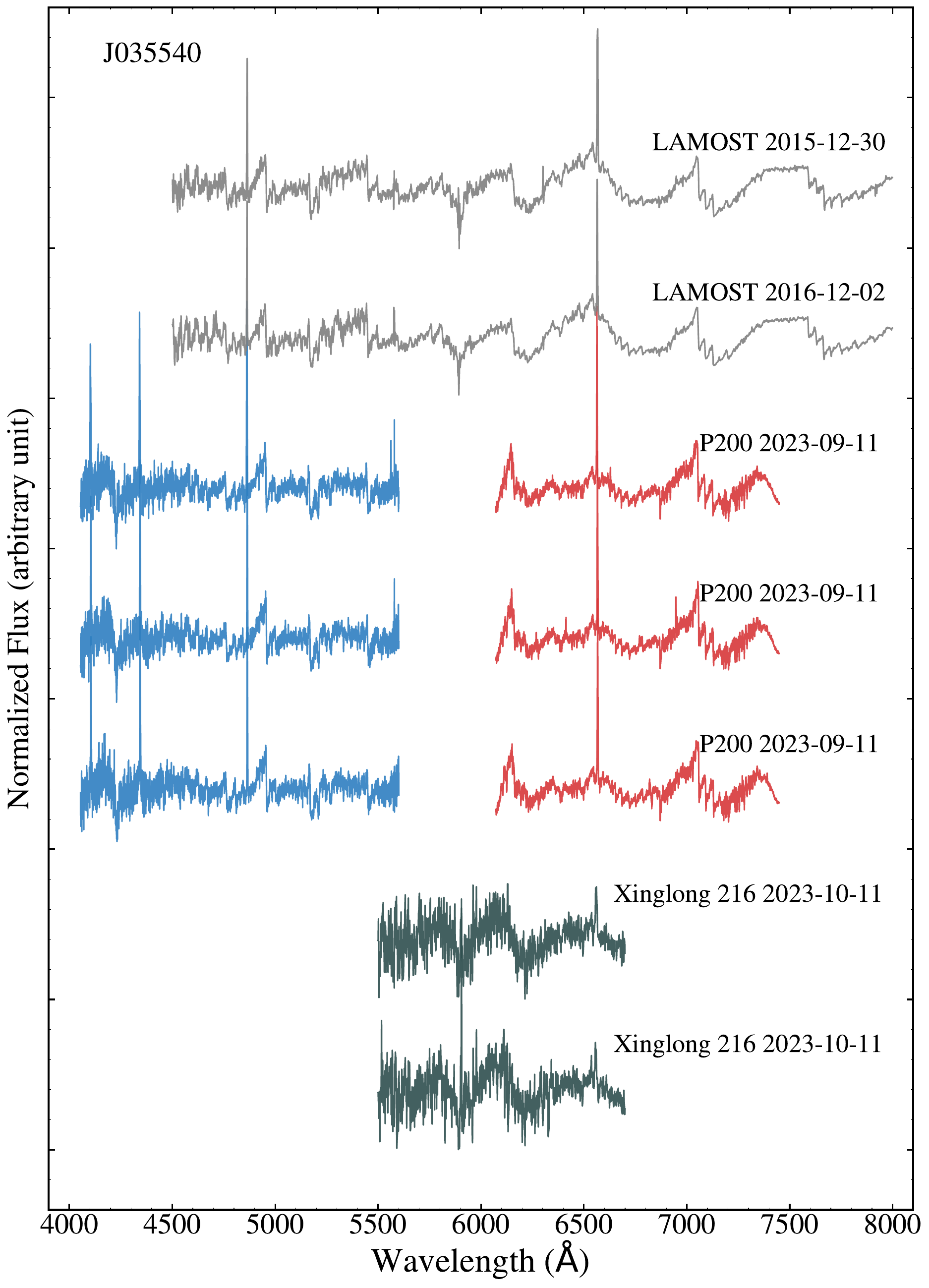}
\includegraphics[scale=0.28]{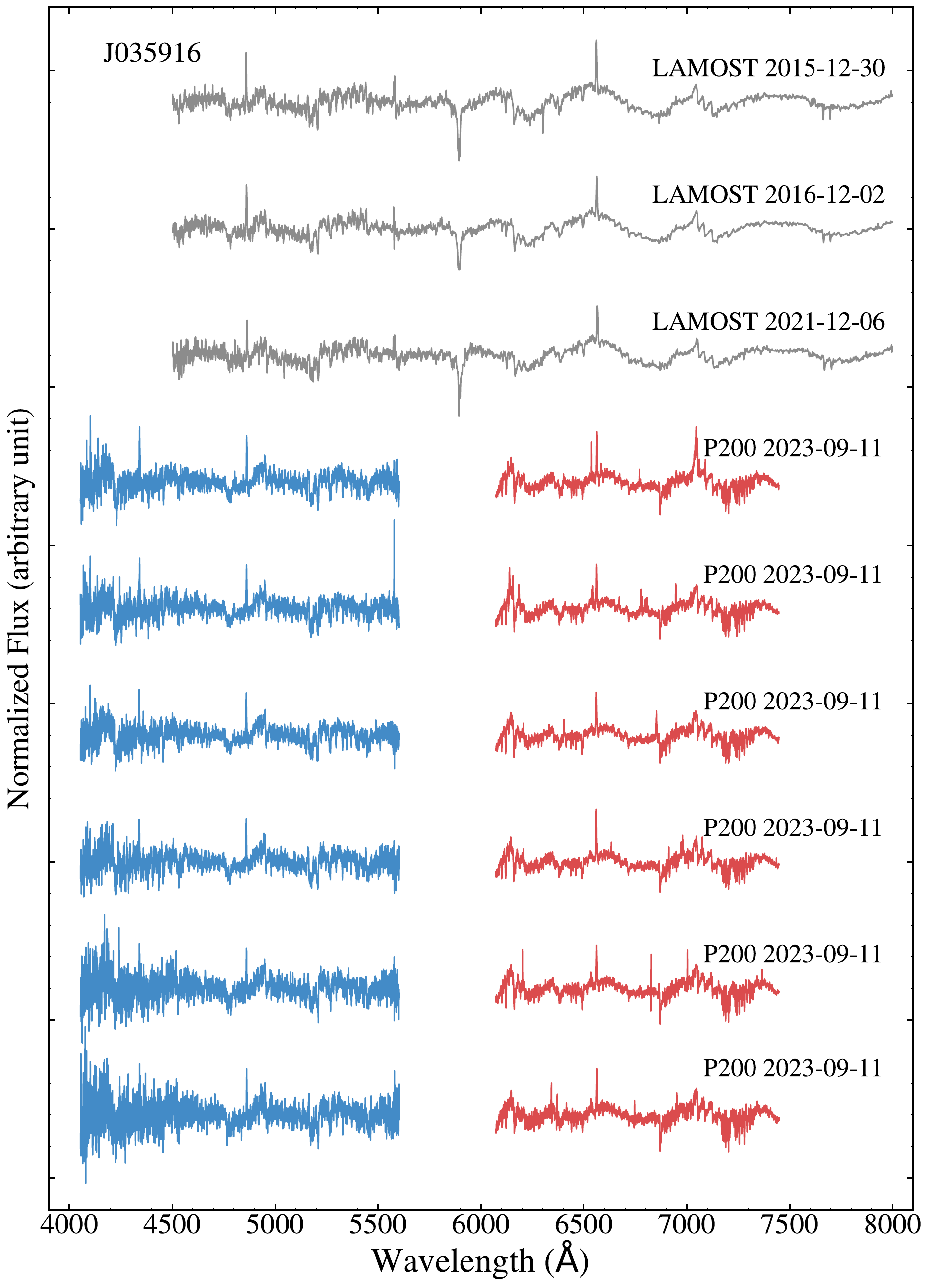}
\caption{Normalized spectra for J035540 and J035916. Observational information is annotated next to each spectrum. LAMOST LRS spectra observed on the same night have been combined while P200 and Xinglong 216 spectra are single exposures.}
\label{fig:fig1}
\end{figure*}

The combined LAMOST spectra for J035540 and J035916 are displayed in Figure~\ref{fig:fig1} (gray). For J035540, LAMOST LRS conducted six observations on December 30, 2015, and December 2, 2016, each with three consecutive 1800-second exposures. J035540 exhibited distinct radial velocity variations on both observation nights. On December 30, 2015, a rapid change of approximately $\mathrm{88\  km \ s^{-1}}$ was observed within one hour, while on December 2, 2016, a similar velocity change of about $\mathrm{84\  km \ s^{-1}}$ occurred within a similar timeframe. 

In the case of J035916, LAMOST LRS made a total of nine observations. These observations were scheduled on December 30, 2015, December 2, 2016, and December 6, 2021, with three consecutive exposures. Each exposure lasted 1800 seconds. Notably, the most significant change in radial velocity occurred during the observation on December 2, 2016, transitioning rapidly from $\mathrm{-127 \  km \ s^{-1}}$ to $\mathrm{-14.6\  km \ s^{-1}}$ within one hour.
Detailed data are presented in Table \ref{table:1}.

\subsection{The P200 spectra} 

We conducted follow-up observations on J035916 and J035540 using the Double Spectrograph (DBSP) mounted on the Palomar Observatory's 200-inch telescope (P200). The observations were taken during a dark night on September 11, 2023. The DBSP was configured with a wavelength range of 3900-5400 \r{A}, achieving a resolving power R$\sim$3400 for the blue arm, and a wavelength range of 6300-7900 \r{A}, with R$\sim$4500 for the red arm. Wavelength calibration for the blue and red arms involved the utilization of HeNeAr and FeAr lamps, respectively.

For J035916, we took a total of 6 spectra. Except for the initial spectrum with a 1200-second exposure, the remaining spectra were exposed for 1500 seconds each. As for J035540, we obtained a total of 3 spectra, each with an exposure time of 1200 seconds. Observation details are listed in Table~\ref{table:1}.

We used the Python package \texttt{Pypeit}\footnote{\url{https://pypeit.readthedocs.io/en/latest/}} \citep{pypeit:joss_arXiv} to reduce the spectra in the blue arm. Since \texttt{Pypeit} does not support the wavelength calibration of the red arm spectrum of the DBSP 1200/7100 D55, we turned to the \texttt{IRAF} to process the spectra in the red arm. The heliocentric correction was performed using the \texttt{pyasl} package \texttt{PyAstronomy}\footnote{\url{https://github.com/sczesla/PyAstronomy}} \citep{pya}. The resulting reduced spectra are depicted in Figure \ref{fig:fig1}.

\subsection{The Xinglong 2.16m Telescope Spectra}

We applied for spectral observations of J035916 and J035540 using the 2.16 m telescope situated at the Xinglong Observatory on October 12, 2023. We chose the G7 grism paired with a 1.8 arcsec slit configuration. 
We exposed both targets to two consecutive observations, each lasting 1800 seconds. We used the \texttt{IRAF} to reduce the spectra using standard procedures. The heliocentric correction was performed using the \texttt{PyAstronomy}.

Due to the extreme faintness of both target sources and adverse weather conditions on the nights of observation, the obtained spectral quality was exceptionally poor. The spectra of J035540 exhibited an extremely low signal-to-noise ratio, rendering the measurement of radial velocity unfeasible. However, two spectra of J035540 did reveal a clear $\rm{H}{\alpha}$ emission line (see Figure \ref{fig:fig1}). In contrast, the spectra of J035916 did not exhibit any distinct spectral structures due to the low quality, thus are not presented in Table \ref{table:1}.

\section{results}
\label{sec:result}

\subsection{The Gaia information}

We cross-match the two sources with Gaia data release 3 \citep[Gaia DR3;][]{2023A&A...674A...1G} using a matching radius of 2\arcsec. 
J035540's Gaia DR3 ID is 220660614918463232, and the \texttt{phot\underline{~}g\underline{~}mean\underline{~}mag} is recorded at 16.37 mag. Gaia DR3 provides a parallax of $\varpi = 4.64\pm0.06$ mas, corresponding to a distance of $d = 215.5\pm2.8$ pc.
For J035916, its Gaia DR3 ID is 229858064046205312. with \texttt{phot\underline{~}g\underline{~}mean\underline{~}mag} of 16.75 mag. Gaia DR3 provides a parallax measurement of $\varpi = 2.90\pm0.06$ mas, inverting to a distance of $d = 344.8\pm7.0$ pc. Astrometric measurements by Gaia are listed in Table~\ref{table:2}.

\subsection{Radial Velocity Measurements}

\begin{figure*}
\centering
\includegraphics[scale=0.60]{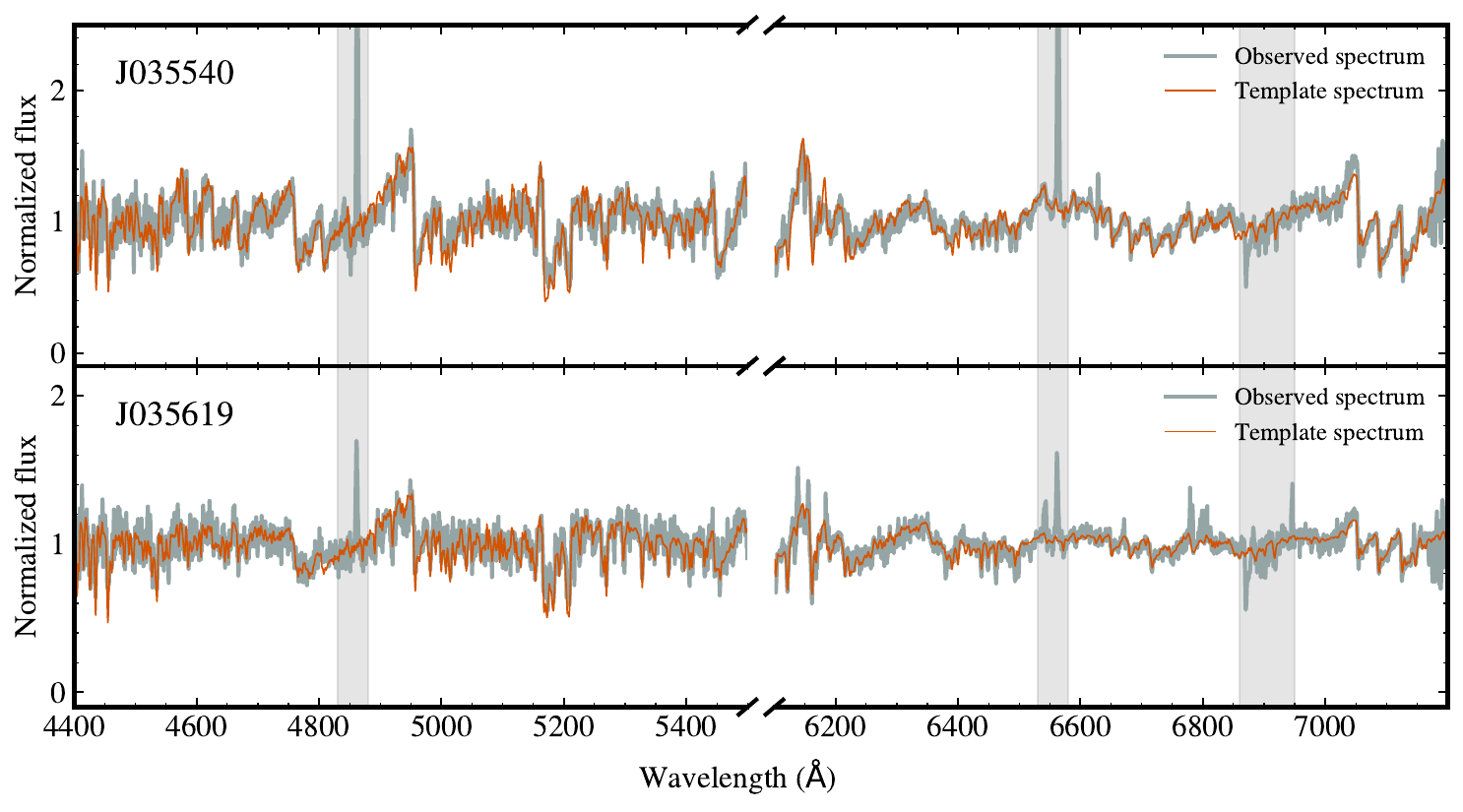}
\caption{The orange curves represent the best-matching model spectra used for radial velocity measurements. The grey curves represent the P200 spectrum for the blue and red arms. The grey shading indicates the wavelength of the observed spectrum contaminated by emission or telluric lines, which is masked during the fitting process.}
\label{fig:fig2}
\end{figure*}

To measure radial velocities, we employ the MARCS template, which is a set of theoretical stellar model atmospheres and flux sample files based on the Uppsala software by the same name \citep{1975A&A....42..407G,2008A&A...486..951G}. The optimal matching spectral template is selected by interpolating the MARCS model grid, minimizing the chi-square between the observation and the template. Figure \ref{fig:fig2} illustrates the observed spectrum and the best-matched template spectrum. Both the blue and red sides of LAMOST LRS and P200 spectra are used to measure the radial velocities, using wavelength ranges of 4500-5600 \r{A} (blue) and 6000-8700 \r{A} (red). Regions containing emission lines and strong telluric absorption lines are masked.

We use the Cross-Correlation Function (CCF) method to measure the radial velocity by cross-matching the template spectrum with the observed spectrum. Simultaneously, we verify that two systems are single-lined spectroscopic binaries by confirming the single-peaked nature of the CCF profile \citep{2017A&A...608A..95M,2021ApJS..256...31L}.
The radial velocity uncertainties are estimated using the flux randomization/random subset sampling (FR/RSS) method, employing a Monte Carlo approach. This involves randomly generating a new flux point within the error range for each spectrum flux point to create a new spectrum with random removal of some data points. This Monte Carlo process is repeated 1000 times, and the standard deviation of the resulting mock radial velocity distribution is adopted as the velocity uncertainty. Table \ref{table:1} summarizes the measured radial velocities and their uncertainties.


\subsection{Orbital parameters}
\label{subsec:3.3}

We apply the Lomb-Scargle algorithm \citep{1976Ap&SS..39..447L,1981ApJS...45....1S} to search for orbital periods ($P_{\rm{orb}}$) in the radial velocity data points of J035540 and J035916. However, the sparse phase coverage of the radial velocity data makes these results less reliable. To obtain more robust period information, we turn to their photometric data. We collect the Zwicky Transient Facility \cite[ZTF;][]{2019PASP..131a8002B} DR19 data for both targets. For J035540, its ZTF g-band photometric dataset has 406 points during the observation from 2018 July 2 to 2023 February 16. The r-band has 851 points during the observation from 2018 July 13 to 2023 July 3. Similarly, J035916's ZTF g-band and r-band photometric datasets consist of 386 and 810 data points, spanning from 2018 March 27 to 2023 April 2, and July 21, 2018, to July 3, 2023, respectively. The Lomb-Scargle method is then employed on the photometric data points to search for their respective photometric periods. 

In the case of J035540, the corresponding peak period does not fit well with the radial velocity curve. Subsequently, we employ the period corresponding to the second-highest peak in the periodogram to fit both the radial velocity curve and the light curve. Remarkably, twice this period effectively fits the variations in both curves. This phenomenon is prevalent in ellipsoidal variables, where the light curve exhibits two peaks and valleys due to tidal deformation of the visible companion under gravitational forces. This results in a quasi-sinusoidal flux variation caused by changes in surface area. Consequently, we deduce the orbital period as 0.4768791 days.
For J035916, fitting the highest power peak in the periodogram to both the light curve and radial velocity data yields an orbital period of 0.3904818 days.

\begin{figure*}
\centering
\includegraphics[scale=0.52]{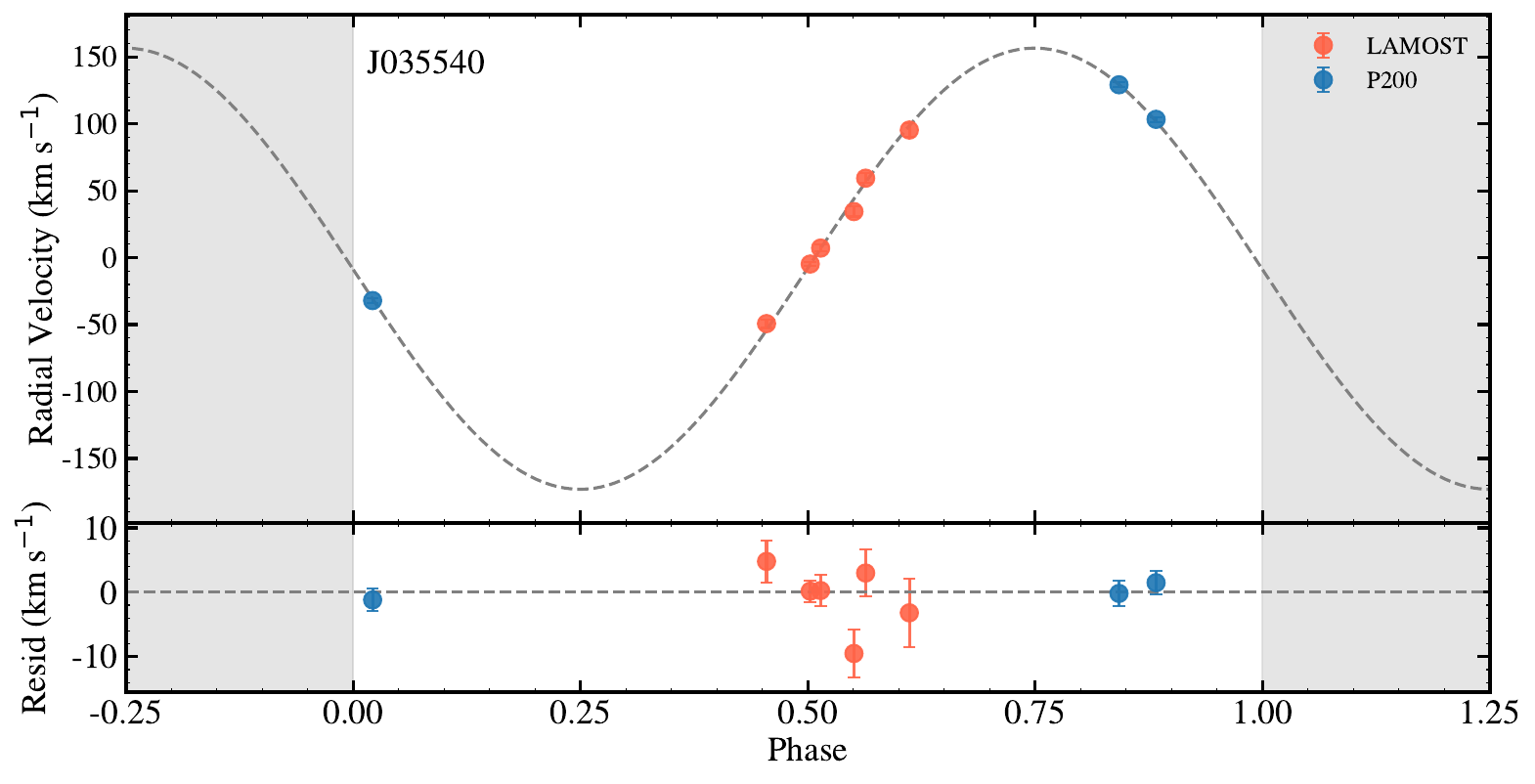}
\includegraphics[scale=0.52]{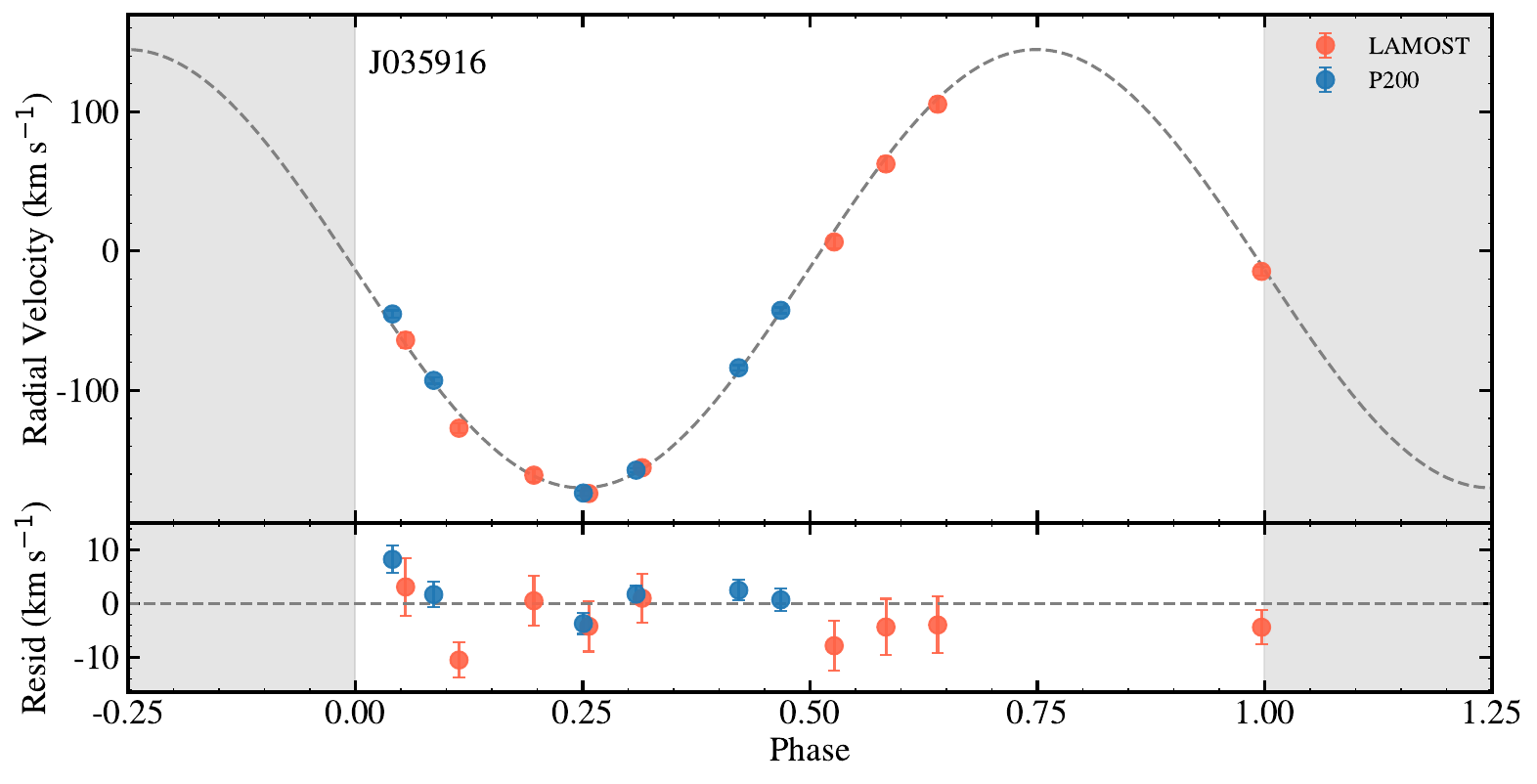}
\caption{The fitted radial velocity curves of J035540 and J035916. The dots represent the radial velocities from LAMOST and P200, while dashed lines depict the best-fitting radial velocity curves.}
\label{fig:fig3}
\end{figure*}

We assume a circular orbit for the binary system and utilize the sinusoidal curve $V_{\rm{R}}=-K_1 \sin[2\pi (t-T_0)/P_{\rm{orb}}]+\gamma$ to fit the radial velocities, where $K_1$ is the radial velocity semi-amplitude, t is the mid-time of each exposure, $T_0$ is the epoch of superior conjunction, and $\gamma$ is the systemic velocity. 
We use the radial velocities measured from the red arm of the spectrum for fitting, since the red arm spectra exhibits higher SNR and resolution than the blue one.
For J035540, the fitting results indicate that $K_1=164.8\pm2.1 \ \mathrm{km\,s^{-1}}$ and $\gamma=-8.3\pm1.0 \ \mathrm{km\,s^{-1}}$, while for J035916, $K_1=159.4\pm1.7 \ \mathrm{km\,s^{-1}}$ and $\gamma=-9.3\pm1.2 \ \mathrm{km\,s^{-1}}$. The fitted radial velocity curves are shown in Figure \ref{fig:fig3} and the fitted orbital parameters are listed in Table \ref{table:2}.

\subsection{Stellar Parameters}

For J035540 and J035916, we conduct broadband spectral energy distribution (SED) fitting to constrain the stellar parameters. We use the python package \texttt{astroARIADNE}\footnote{\url{https://github.com/jvines/astroARIADNE}}, which employs the Nested Sampling algorithm to automatically fit the SED of target stars \citep{2022MNRAS.513.2719V}.

\begin{figure}
  \centering
  \begin{minipage}[b]{0.43\textwidth}
    \centering
    \includegraphics[width=\textwidth]{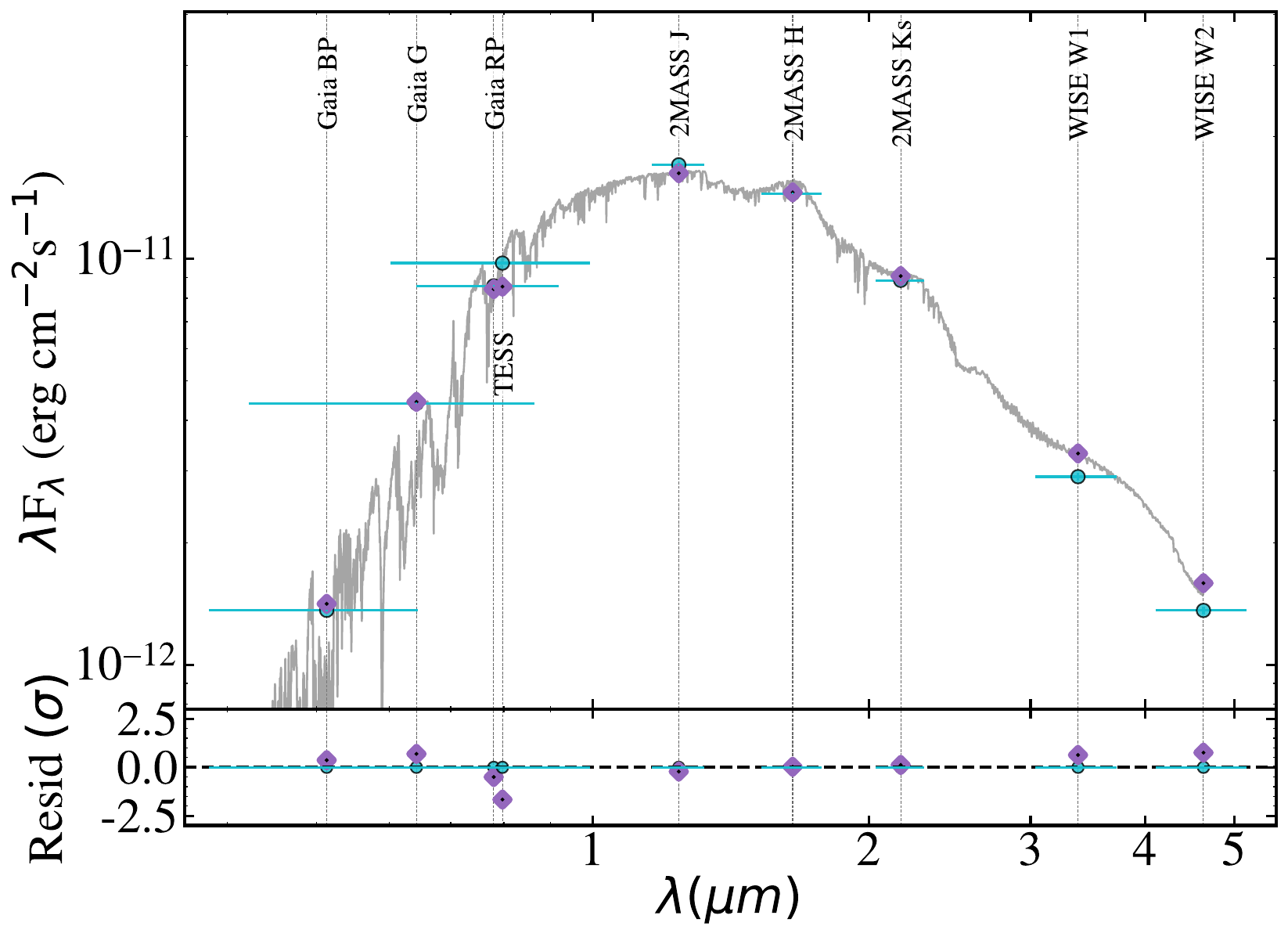}
  \end{minipage}
  \hfill
  \begin{minipage}[b]{0.43\textwidth}
    \centering
    \includegraphics[width=\textwidth]{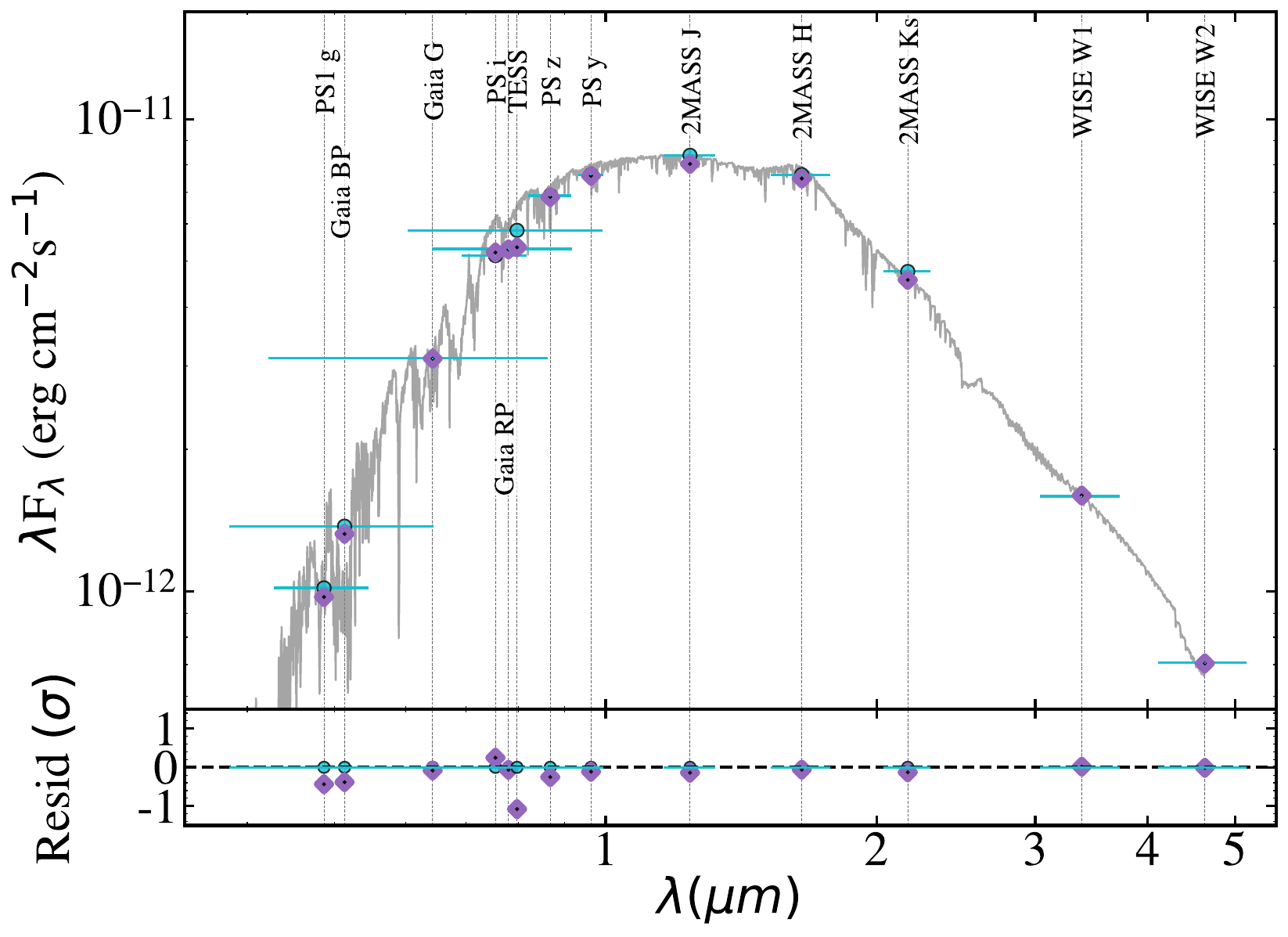}
  \end{minipage}
  \caption{SED fitting results for J035540 (top panel) and J035916 (bottom panel). Cyan points denote observed fluxes, purple diamonds indicate synthetic fluxes, and the grey curve represents the best-fitting SED model.}
  \label{fig:fig4}
\end{figure}

The SED fitting procedure involves the following steps. Firstly, \texttt{astroARIADNE} uses the python package \texttt{astroquery} to automatically collect multi-band photometric data, including GALEX \citep{2005ApJ...619L...1M,2011Ap&SS.335..161B}, SDSS \citep{2009ApJS..182..543A}, APASS \citep{2014CoSka..43..518H}, Pan-STARRS \citep{2016arXiv161205560C}, TESS \citep{2015JATIS...1a4003R}, 2MASS \citep{2006AJ....131.1163S}, Kepler \citep{2010Sci...327..977B}, NGTS \citep{2018MNRAS.475.4476W}, and ALLWISE \citep{2010AJ....140.1868W}. 
Subsequently, parallax data from Gaia DR3 are used as prior for the SED fitting distance.
\texttt{astroARIADNE} uses six distinct atmospheric model grids: PHOENIX v2 \citep{2013A&A...553A...6H}, Castelli \& Kurucz \citep{2003IAUS..210P.A20C}, BT-Settl \citep{2012RSPTA.370.2765A}, BT-Cond \citep{2012RSPTA.370.2765A}, BT-NextGen \citep{1999ApJ...525..871H,2012RSPTA.370.2765A}, and Kurucz \citep{1993KurCD..13.....K} to fit the SED and derive stellar parameters such as effective temperature, surface gravity, metallicity, distance, radius, and V-band extinction. The resulting best-fitting stellar parameters are summarized in Table \ref{table:2}, and the corresponding models are illustrated in Figure \ref{fig:fig4}.

After the SED fitting, we estimate the masses of the visible stars in J035540 and J035916 using stellar evolution models. The SED best-fitting parameters and photometry are served as inputs for deriving isochrone masses with the MESA Isochrones \& Stellar Tracks \citep[MIST;][]{2016ApJS..222....8D} isochrones. This procedure is implemented in the python package \texttt{isochrones}\footnote{\url{https://isochrones.readthedocs.io/en/latest/}}. The isochrone masses are presented in Table \ref{table:2}.

\subsection{Mass constraints}

Using the the radial velocity semi-amplitude $K_1$ for the visible stars and $P_{\rm{orb}}$, the mass function for the invisible companion can be calculated as:
\begin{linenomath}
\begin{equation}\label{eq1}
f(M_2) \equiv \frac{M_2^{3}\sin^{3}i}{{\left(M_1+M_2\right)}^2} = \frac{K^3_1P_{\rm{orb}}}{2\pi G},
\end{equation}
\end{linenomath}
where $M_1$ is the mass of the visible star and $M_2$ is the mass of the invisible companion, $i$ is the orbital inclination angle, and \textit{G} is the gravitational constant.
The mass function provides a robust lower limit on the mass of the unseen companion when $i=90^{\circ}$ and $M_1 = 0 M_{\odot}$.
The mass functions for J035540 and J035916 are $0.22 \pm 0.01 M_{\odot}$ and $0.16 \pm 0.01 M_{\odot}$, respectively.

We adopt the isochrone mass as the mass estimate for the visible stars. Utilizing the previously determined mass functions for the unseen companions, we calculate the lower mass limit for these invisible stars, assuming a binary orbital inclination of 90 degrees. The $M^{\rm{min}}_2$ of the unseen companion in J035540 is $0.64 \pm 0.04 \ M_{\odot}$, while for J035916, it is $0.57 \pm 0.05 \  M_{\odot}$.

The results indicate that the lower mass limit of the unseen companions in both J035540 and J035916 exceeds the mass of the visible stars. In typical binary systems with similar masses, the companion stars contribute significantly to the overall flux and exhibit distinct double-lined spectroscopic features, which are not observed in the obtained spectra. Therefore, it can be concluded that both systems harbor hidden compact stars.


\begin{table*}
\caption{Summary information for J035540 and J035916, including astronomical parameters from Gaia DR3, orbital parameters derived from ZTF light curves and radial velocities measured from LAMOST LRS and P200 spectroscopic data, and stellar parameters determined through SED fitting.}
\label{table:2}
\begin{adjustbox}{width=1\textwidth}
\begin{tabular}{llrrl}
\hline\hline
Parameter & Unit & J035540 & J035916 & Note\\
\hline
Astrometrical parameters &  &  &  &  \\
R.A. & h:m:s(J2000) & 03:55:41 & 03:59:16 & R.A. \\
Decl. & d:m:s(J2000) & +38:15:50.0 & +40:07:32.16 & Decl. \\
$\pi$ & mas & 4.64 $\pm$ 0.06 & 2.90 $\pm$ 0.06 & Parallax from Gaia DR3 \\
$d$ & pc & 215.5 $\pm$ 2.8 & 344.8 $\pm$ 7.0 & Distance estimated using Gaia DR3 parallax  \\
${\mu}_{\alpha}$ & $\mathrm{mas\,yr^{-1}}$ & -1.20 $\pm$ 0.07 & -5.61 $\pm$ 0.08 & Proper motion in R.A. direction from Gaia DR3  \\
${\mu}_{\delta}$ & $\mathrm{mas\,yr^{-1}}$ & -50.61 $\pm$ 0.05 & -0.74 $\pm$ 0.06 & Proper motion in Decl. direction from Gaia DR3  \\
G-band magnitude & mag & 16.37 $\pm$ 0.01 & 16.75 $\pm$ 0.01 & G-band magnitude from Gaia DR3  \\ \hline
Orbital parameters &  &  &  &  \\
$P_{\rm{orb}}$ & days & 0.4768791 & 0.3904818 & Orbital period \\
$K_1$ & $\mathrm{km\,s^{-1}}$ & 164.8 $\pm$ 2.0 & 157.2 $\pm$ 1.7 & RV semi-amplitude of the visible star \\
$\gamma$ & $\mathrm{km\,s^{-1}}$ & -8.3 $\pm$ 1.0 & -12.7 $\pm$ 1.2 & Systemic velocity \\
$f(M_2)$ & $M_{\odot}$ & 0.22 $\pm$ 0.01 & 0.16 $\pm$ 0.01 & Mass function of the unseen star \\ \hline
Stellar parameters of the visible stars &  &  &  &  \\
$T_{\rm{eff}}$ & K & $3304^{+99}_{-81}$ & $3672^{+102}_{-87}$ & Effective temperature measured by SED fitting \\
log$g$ & dex & $4.98^{+0.23}_{-0.24}$ & $4.75^{+0.04}_{-0.04}$ & Surface gravity measured by SED fitting \\
$M_1$ & $M_{\odot}$ & $0.46\pm0.04$ & $0.51\pm0.05$ & Mass of visible star from the MIST model  \\
$R_1$ & $R_{\odot}$ & $0.57^{+0.04}_{-0.05}$ & $0.54^{+0.02}_{-0.03}$ & Radius measured by SED fitting \\

\hline
\end{tabular}
\end{adjustbox}
\end{table*}


\section{discussion}
\label{sec:discussion}

\subsection{Light curve and filling factor}\label{subsec:4.1}

Due to the relatively faint nature of J035540 and J035916, the photometric data from ZTF exhibit significant uncertainties. Nevertheless, distinctive variations in the photometric flux are visually apparent in the noisy phased light curves. The folded light curves reveal the characteristic signature of ellipsoidal modulated variability, as discussed in Section \ref{subsec:3.3}. The standard ellipsoidal modulation light curve features a double-peak-double-valley shape, a consequence of tidal distortion of the star by a companion. The star is elongated to an ellipsoidal shape. Furthermore, gravity-darkening induces different minima at the two distinct peaks when the longest axis of the ellipsoidal body is oriented towards the observer \citep{2021MNRAS.508.4106E}. However, this second-order effect is nearly imperceptible in the noisy light curve.

\begin{figure*}
\centering
\includegraphics[scale=0.35]{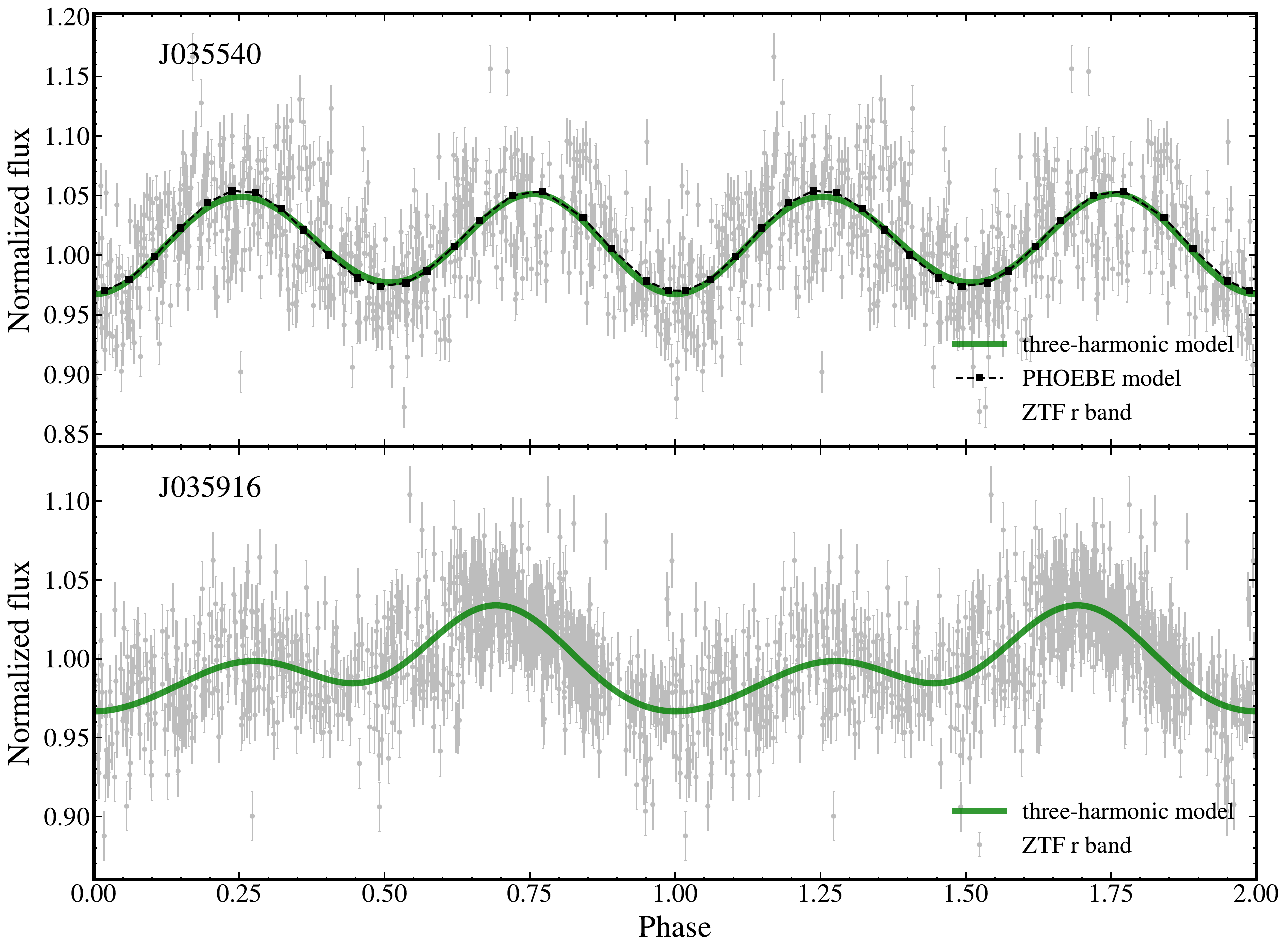}
\caption{ZTF light curves (grey dots) of J035540 and J035916 folded by their respective orbital periods. The green curves depict the three-harmonic model based on Equation \ref{eq2}. In the top panel, the black dotted line represents the best-fitting \texttt{PHOEBE} model for J035540.}
\label{fig:fig5}
\end{figure*}

To enhance the clarity of the light curve profile, we use a three-harmonic model for fitting the normalized light curve. This model can be expressed as \citep{1993ApJ...419..344M}:
\begin{linenomath}
\begin{equation}
\label{eq2}
\begin{split}
f(t) =\ & {a_1}\cos[\frac{2\pi}{P_{\rm{orb}}}(t-{T_0})]+{a_2}\cos[\frac{4\pi}{P_{\rm{orb}}}(t-{T_0})]\\
        &+{a_3}\cos[\frac{6\pi}{P_{\rm{orb}}}(t-{T_0})],
\end{split}
\end{equation}
\end{linenomath}
where $a_1$, $a_2$, and $a_3$ are the fitted parameters.
The fitted model reveals that the light curve of J035540 exhibits a standard ellipsoidal modulation, while the light curve of J035916 displays an asymmetric double-peaked variation.
The light curves and the fitting model for two targets are shown in Figure \ref{fig:fig5}

After determining their binary orbital parameters and stellar parameters of the visible star, we calculate their Roche lobe filling factor. The Roche lobe filling factor is defined as $f \equiv R_1/R_{\rm{L1}}$, where $R_{\rm{L1}}$ is the Roche lobe radius of the visible star. 
We calculate $R_{\rm{L1}}$ using the equation (4) of \citet{2023AJ....165..187Q}: 
\begin{linenomath}
\begin{equation}\label{eq3}
\frac{M_{1}}{R^{3}_{\rm{L1}}} = 0.804 \ P^{-2}_{\rm{day}}\ {\rm g\,cm^{-3}},
\end{equation}
\end{linenomath}
where $P_{\rm{day}}$ is the orbital period in the unit of days.
We use Equation~(\ref{eq3}) to estimate $R_{\rm{L1}}$, while $R_{1}$ is measured from SED fitting. Following this, the filling factors for J035540 and J035916 are calculated as $0.62 \pm 0.06$ and $0.65 \pm 0.05$, respectively, aligning with the observed significant variations in the light curve.

\subsection{Constraining the invisible object's mass}\label{subsec:4.2}

Thus far, we have established the lower bounds for the masses of the compact objects in J035540 and J035916 using the mass function, assuming an orbital inclination of $i = 90 ^{\circ}$. However, to determine the actual masses, we need to establish the orbital inclination of the binary systems.

J035540 displays a characteristic double-peaked ellipsoidal modulation, with key influencing factors being the binary inclination $i$, mass ratio $q = M_{\rm{2}}/M_{\rm{1}}$, Roche lobe filling factor $f$, limb-darkening coefficient, and gravity-darkening coefficient. To determine the values of $i$ and $q$, we employed the \texttt{PHOEBE} software \citep{2005ApJ...628..426P,2011ascl.soft06002P,2020ApJS..250...34C} for light curve fitting. 

During the light curve fitting, we set \texttt{distortion method $=$ none}, treating the invisible companion as an object with no flux or eclipse contribution. Simultaneously, we set the orbital period and effective temperature of the visible star to the values listed in Table \ref{table:2}. Constraints on the projection of the orbital semi-major axis are applied using the orbital period and radial velocity semi-amplitude. The limb-darkening coefficient is manually set, following a logarithmic limb-darkening law, with values of [0.8, 0.3] \citep[see][]{2011A&A...529A..75C}. The gravity-darkening coefficient is treated as a free parameter, with a prior value of $\beta_1 \sim \mathcal{N}(0.5, 0.2)$ \citep[see][]{2011A&A...529A..75C}. We employ the \texttt{emcee} package for Markov Chain Monte Carlo (MCMC) fitting, using the provided values for radius, mass, and distance in Table \ref{table:2} as priors. Multiple parallel chains are run, each consisting of 10,000 steps.

\begin{figure}
\centering
\includegraphics[width=0.44\textwidth]{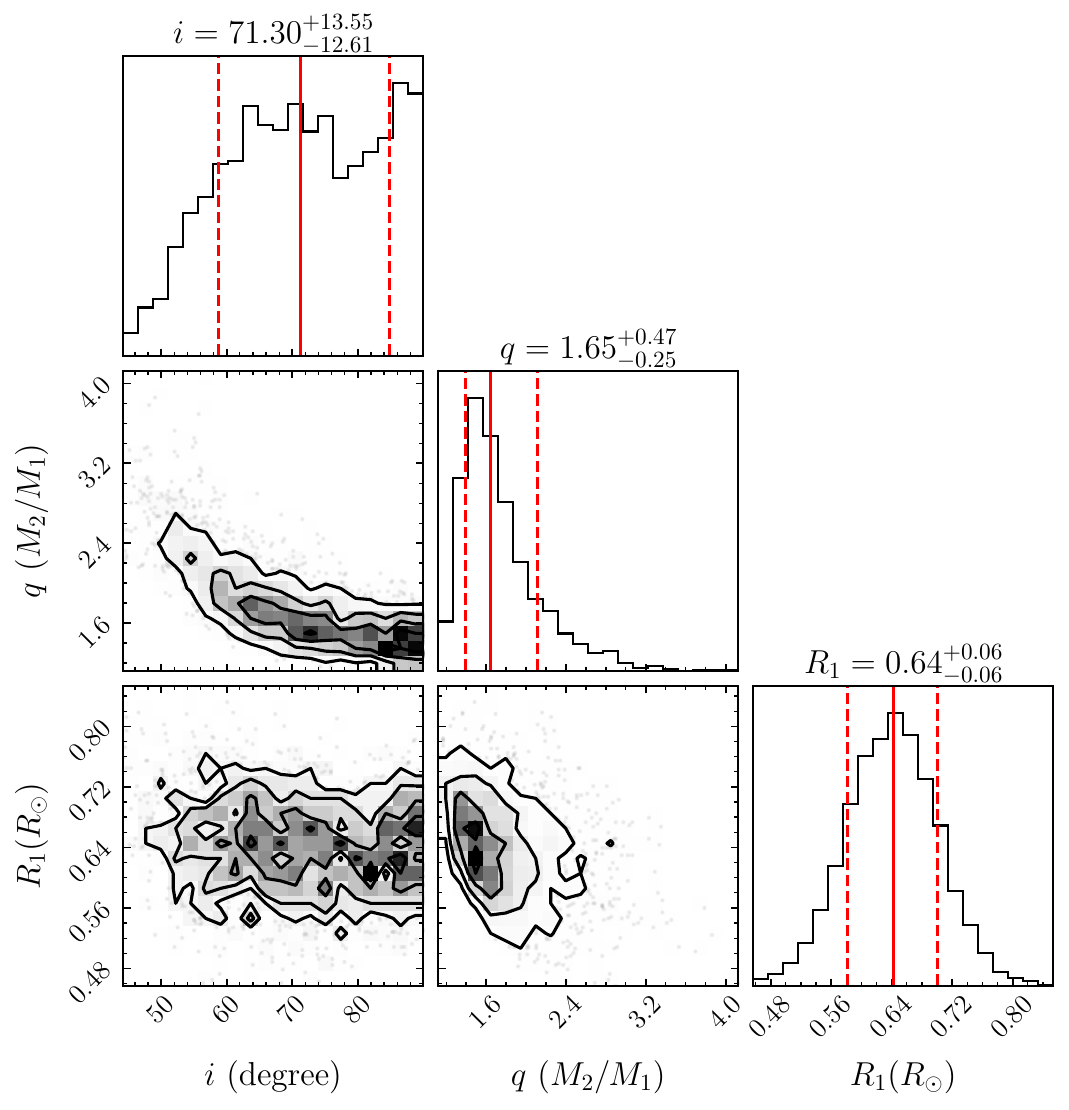}
\caption{Parameter distribution from the \texttt{PHOEBE} light curve fitting of J035540.}
\label{fig:fig6}
\end{figure}

In Figure \ref{fig:fig6}, the posterior samples for $q$, $i$, and $R_1$ are presented. The fitting results indicate $i=71.30^{+13.55}_{-12.61}$ degrees for J035540, along with an associated mass for the unseen companion of $M_2=0.70^{+0.12}_{-0.05} M_\odot$. The optimal \texttt{PHOEBE} model outcomes are represented as the black lines in Figure \ref{fig:fig5}, indicating that the ellipsoidal model can fit the observed light curve aptly.

The light curve of J035619 exhibits an asymmetric double-peaked structure, possibly attributed to spots on the visible star's surface \citep{2021AJ....162..123L,2024MNRAS.529..587R}. Modeling this asymmetric light curve with \texttt{PHOEBE} is highly dependent on the spot model.
In our analysis, we use the work proposed by \cite{2021MNRAS.501.2822G} to fit the light curve of J035916 for providing certain constraints on the inclination. This approach is derived from the work of Morris \& Naftilan 1993 (MN93) \citep{1993ApJ...419..344M} and offers an analytical approximation for ellipsoidal modulation under the assumption of a circular orbit.
The leading corrected approximation can be written as
\begin{linenomath}
\begin{equation}\label{eq4}
\frac{{\Delta}L}{\overline{L}} \  \approx  \ \frac{1}{\overline{L}/L_0} \  \alpha_2 \  f^3 \ E^3(q) \ q \ \rm{sin}^2 \textit{i} \ \textit{C(q,\ f)} \ cos2\phi, 
\end{equation}
\end{linenomath}
where $q$ is the mass ratio $q \equiv M_2/M_1$, $\overline{L}$ is the average luminosity, $L_0$ is the stellar brightness without secondary, $\phi$ is the orbital phase and the $\alpha_2$ is a function of the linear limb- and gravity-darkening coefficients of the detected star, the $E(q)$ is the \cite{1983ApJ...268..368E} approximation for the volume-averaged Roche lobe radius and $C(q,f)$ is the correction factor related to $q$ and $f$.
The correction factor begins at 1 for $f=0$ (no correction) and rises monotonically to a maximum value of $\sim 1.5$ at $f\gtrsim 0.90$ \citep{2021MNRAS.501.2822G}.
The modified formula demonstrates that the ellipsoidal amplitude depends on the filling factor, the inclination, and the mass ratio. 
The mass function is invoked to place an additional constraint between the inclination angle and mass ratio for a given visible star's mass. Based on the known stellar properties and orbital parameters, the observed ellipsoidal variability can provide a combined constraint on the mass of the invisible component of the binary system.

\begin{figure}
\centering
\includegraphics[width=0.45\textwidth]{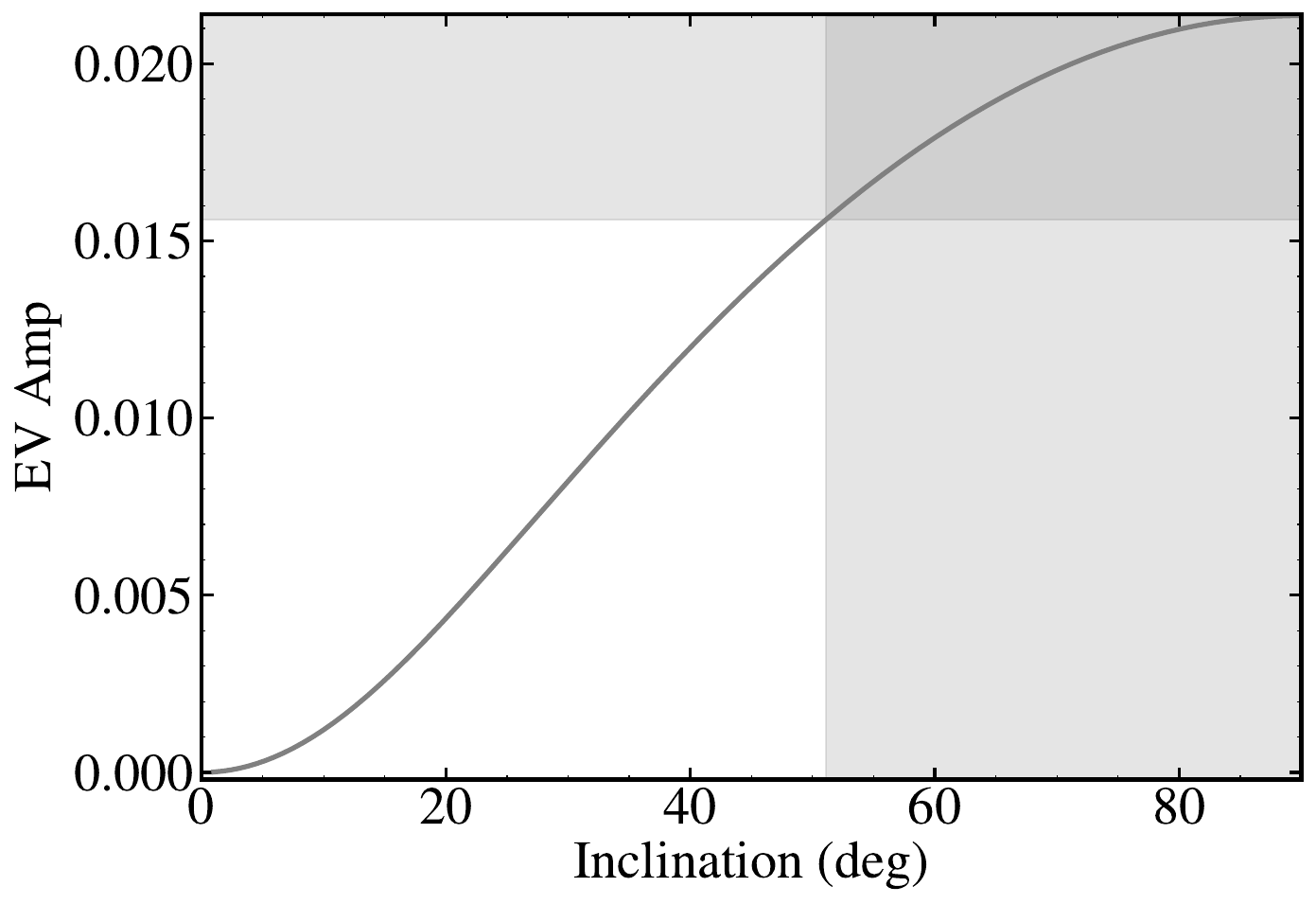}
\caption{The panel depicts the semi-amplitude of ellipsoidal variability versus orbital inclination. The horizontal shaded region represents the ellipsoidal modulation amplitudes observed in J035916's light curve, while the vertical shaded region indicates the corresponding range of orbital inclinations. }
\label{fig:fig7}
\end{figure}

Figure \ref{fig:fig7} illustrates the constraints on the orbital inclination and the mass of the invisible star using the aforementioned analysis model. The graph illustrates the correlation between the half-amplitude of ellipsoidal variations and the orbital inclination. The results reveal that the orbital inclination of J035916 falls within $50^{\circ} - 90^{\circ}$, corresponding to an inferred mass of the compact object ranging from $0.57 - 0.90 \  M_{\odot}$.

We use \texttt{PHOEBE} to confine the inclination of J035540 and subsequently utilize the analysis model to restrict the inclination of J035916. 
However, due to the noisy nature of the photometric data used, our constraints on the true orbital inclination remain relatively broad. To further tighten the constraints on the inclination, one can enhance measurements by measuring the rotational broadening of the visible stars from high-resolution spectra \citep{1994MNRAS.266..137M}.

\subsection{The \texorpdfstring{$v_{\rm{rot}} \sin i$}{} measurement}

As discussed in Section \ref{subsec:4.2}, the ZTF light curve of J035916 exhibits asymmetrical features, preventing us from accurately fitting the orbital inclination using an ellipsoidal model. An alternative approach is to independently constrain the inclination by measuring the projected rotational broadening velocity ($v_{\rm{rot}} \sin i$). We selected the P200 red-side spectrum at HMJD 60198.416755 to measure the $v_{\rm{rot}} \sin i$. This spectrum was taken at the quadrature orbital phase ($\phi \sim 0.25$), therefore the radial velocity smearing effect is minimized. We estimated that the velocity smearing is only around $ 6.4 \ \mathrm{km\,s^{-1}}$ due to the star's orbital motion during the 1500s on-target exposure.

Major stellar spectral-line broadening mechanisms include rotational broadening, macro-turbulence broadening, and instrumental broadening. We determined the instrumental broadening by measuring the FWHM (full width at half maximum) of the lamp spectrum taken during the P200 observation. Specifically, a Gaussian kernel was used to fit the lamp spectrum line profile, yielding the FWHM = 1.4 \r{A} at 6402 \r{A}, which corresponds to a spectral resolution of R $\sim$ 4400, or equivalently, a broadening of $\Delta V_{\rm{ins}} \sim 68 \  \mathrm{km\,s^{-1}}$ in the velocity space. 
The theoretical rotational broadening kernel \citep{Gray2005} is as follows:
\begin{linenomath}
\begin{equation}\label{eq:rbk}
    G(x) = \left\{ 
    \begin{array}{lr}
        \frac{2\left(1-\epsilon\right)\left(1-x^2\right)^{1/2}+\frac{\pi\epsilon}{2}\left(1-x^2\right)}{\pi\left(1-\frac{\epsilon}{3}\right)} & \mathrm{for}\ \mid x \mid < 1 \\
        0 & \mathrm{for}\ \mid x \mid > 1
    \end{array}
    \right.,
\end{equation}
\end{linenomath}
where $x = V/v_{\rm{rot}} \sin i$, and $\epsilon$ is the linear limb-darkening coefficient.
We set the macro-turbulence broadening to be 2 $\mathrm{km\,s^{-1}}$. 

We select the spectral range of 6400-6500 \r{A} for fitting,
where a few relatively strong and distinct absorption lines are present. Additionally, it is free of telluric absorption lines. We generate the spectral template from MARCS model atmospheres using the stellar parameters listed in Table \ref{table:2}. 
The template is first convolved with the rotational broadening kernel (Equation~\ref{eq:rbk}), then convolved with Gaussian instrumental broadening kernel. We obtain the $v_{\rm{rot}} \sin i$ by minimizing the chi-square $\chi^{2} = \sum ((o-m)/err)^2$ between the model and observed spectra,
where $o$ and $m$ are the fluxes of the observation and model, respectively; $err$ is the flux uncertainty, and the summation sign is taken over all the flux points with the spectral range. 

\begin{figure}
\centering
\includegraphics[width=0.45\textwidth]{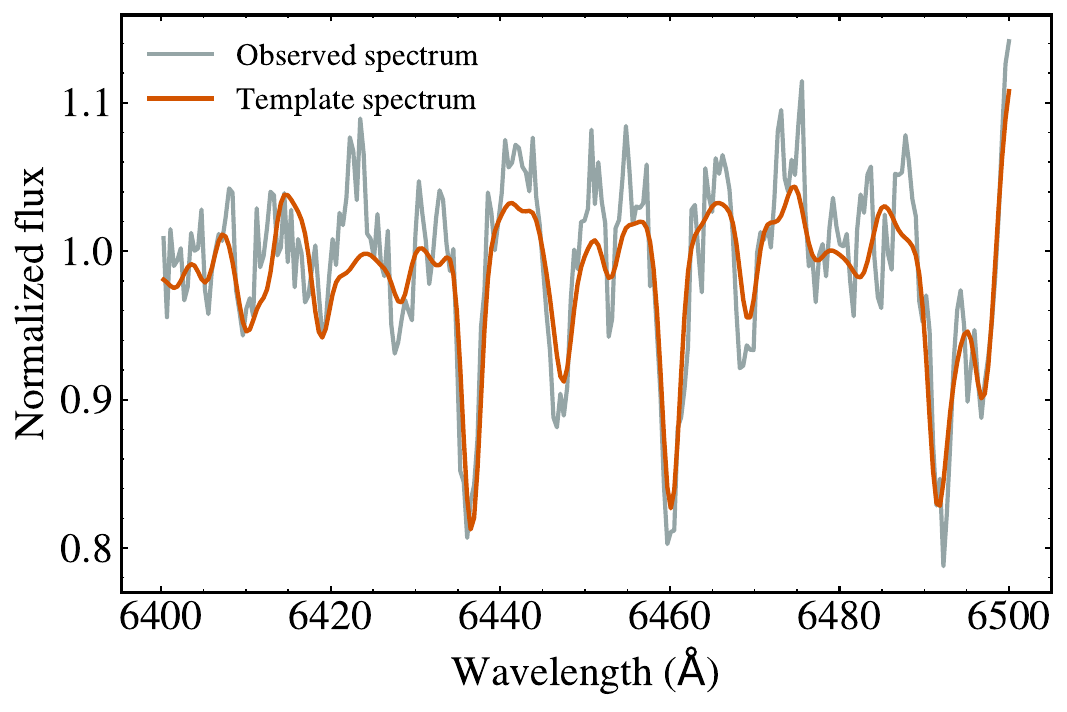}
\caption{Spectral wavelength region used to measure the rotational broadening of J035916. The grey curve represents the observed spectrum and the orange curve represents the rotationally broadened best-fit template.}
\label{fig:fig8}
\end{figure}

The fitting result is $v_{\rm{rot}} \sin i = 59_{-5}^{+6} \  \mathrm{km\,s^{-1}}$. Figure \ref{fig:fig8} presents the optimal fitting outcomes. 
The short orbital period of J035916 implies that the system is tidally synchronized. Consequently, we adopt the assumption that the rotation period ($P_{\rm{rot}}$) of the visible star is equivalent to $P_{\rm{orb}}$. Therefore, we can derive the rotational velocity as $v_{\rm{rot}} = 2 \pi R_{\rm{vis}}/P_{\rm{orb}} = 70 \pm 3 \  \mathrm{km\,s^{-1}}$, where $R_{\rm{vis}}$ is the radius of the visible star. Thus, we estimate that J035916's inclination $i = 57_{-13}^{+18}$ degrees, consistent with that estimated from the analytical ellipsoidal model (Section \ref{subsec:4.2}).

\subsection{Galactic kinematics of two targets}

Based on the astrometrical parameters provided by Gaia DR3 and the systemic radial velocities obtained through spectra, we estimated the Galactic velocities of these two sources relative to the local standard of rest (LSR) using the \texttt{pyasl} package in \texttt{PyAstronomy}. The adopted peculiar velocity of the Sun is $(U_{\rm{LSR}}, V_{\rm{LSR}}, W_{\rm{LSR}}) = (11.1, 12.24, 7.25)\  \mathrm{km\,s^{-1}}$. The Galactic space velocities for J035540 are $(U_{\rm{LSR}}, V_{\rm{LSR}}, W_{\rm{LSR}}) = (13.9\pm0.9, -24.3\pm0.6, -30.1\pm0.6)\  \mathrm{km\,s^{-1}}$, and for J035916 as $(U_{\rm{LSR}}, V_{\rm{LSR}}, W_{\rm{LSR}}) = (26.1\pm1.1, 12.6\pm0.5, 2.6\pm0.3)\  \mathrm{km\,s^{-1}}$. 

\begin{figure}
\centering
\includegraphics[scale=0.45]{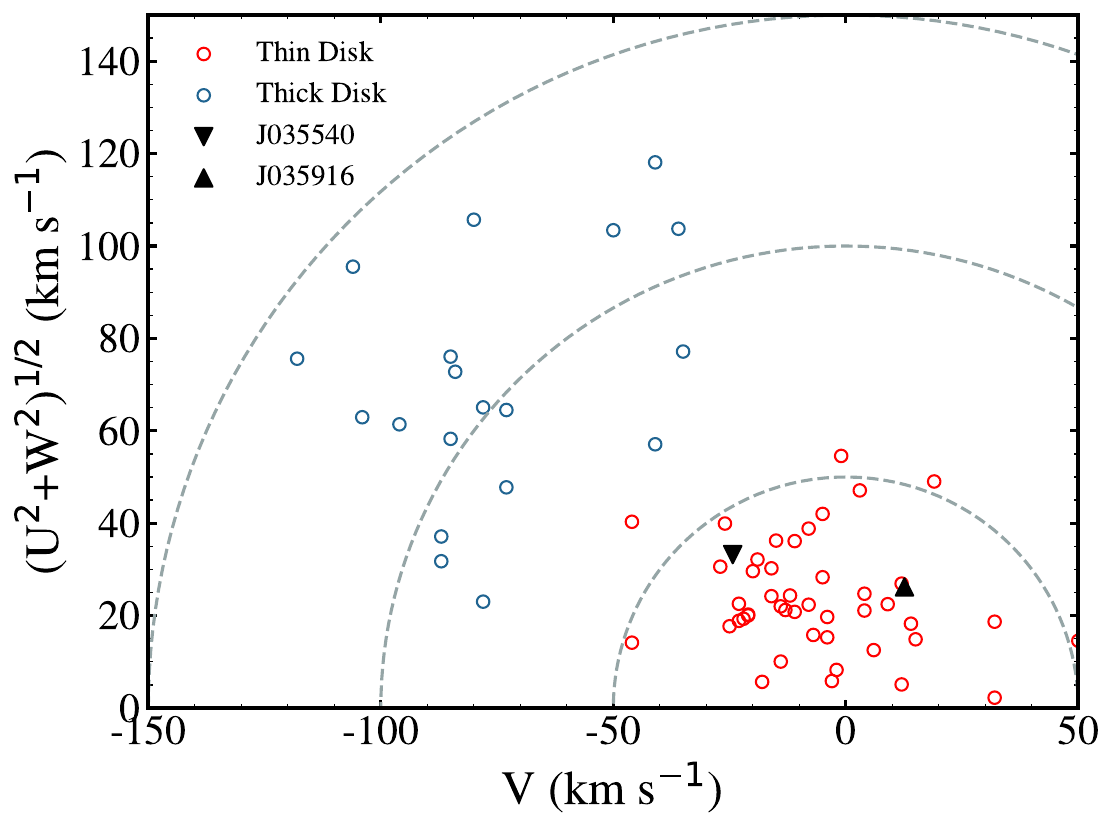}
\caption{Toomre diagram of our sources. Dashed lines indicate constant peculiar space velocities in steps of 50 $\mathrm{km\,s^{-1}}$. Blue dots represent the thick disk stars and red dots represents the thin disk stars adopted from \citet{2003A&A...410..527B}.}
\label{fig:fig9}
\end{figure}

In Figure \ref{fig:fig9}, we depict the locations of two targets in the Toomre diagram, both falling within the region associated with the thin disk. Following this, we compute the thick-to-thin disk probability ratio (TD/D) using the equation (7) from \citet{2017ApJ...850...25L}. The TD/D values are 0.08 for J035540 and 0.02 for J035916, indicating that they are most likely thin disk stars.

\subsection{The nature of two targets}

As described above, we constrain the orbital inclinations of J035540 and J035916 using the light curve, thereby restricting the mass of the unseen stars in these systems. 
The inferred mass of the unseen stars suggests that these two systems are binary systems composed of a WD candidate and an M-type star. The most reliable method to confirm the presence of WDs and to measure their parameters is through their UV spectra \citep{2016MNRAS.463.2125P,2022MNRAS.517.2867H}.
Since a WD has radiation peaked at near UV to UV wavelengths, 
we tried cross-matching J035540 and J035916 with GALEX using a cross-matching radius of 20\arcsec, to search for potential UV counterparts. However, J035540 and J035916 are not covered in GALEX's footprints.

The future Chinese Space Station Telescope (CSST) will conduct a wide-field imaging in the near-ultraviolet-optical-near-infrared (NUV-OP-NIR) bands \citep{article}.
In the context of white dwarfs, we utilize WD cooling models \citep{2020ApJ...901...93B} to assess their detectability by CSST based on assumed temperatures. 
The tables from \url{http://www.astro.umontreal.ca/~bergeron/CoolingModels} provide bolometric and absolute magnitudes for WDs on various photometric systems including NUV colors.
We use the cooling model for a typical DA-type WD with mass of $M_2 = 0.60 M_{\odot}$ to calculate the absolute NUV magnitudes across a grid of WD temperatures.
The visual NUV magnitude from CSST is $m_{\rm{NUV}} = M_{\rm{NUV}} + 5{\rm{log}}(d) + A_{\rm{NUV}} -5$, where $M_{\rm{NUV}}$ is the absolute NUV magnitude obtained from the cooling model, $d$ is the distance from Gaia DR3, and $A_{\rm{NUV}}$ is the NUV extinction value. The $A_{\rm{NUV}}=0.25$ is derived from the Milky Way Average Extinction Curve \citep{2009ApJ...705.1320G}. CSST has a detection limiting magnitude for NUV at approximately 25.4 mag.
We conclude that the compact objects in J035540 and J035916, if they are DA-type WDs, will be detected by CSST when their temperatures exceed 6000K and 6500K, respectively.

We use the dynamical method for the identification of two binary systems that potentially harbor unseen WDs. This method distinguishes itself from both the spectroscopic decomposition method \citep{2014MNRAS.445.1331L,2016MNRAS.458.3808R} and the UV-excess selection method \citep{2016MNRAS.463.2125P}. 
More than ten sources identified through these methods have also undergone the dynamical measurement to further investigate their binary properties \citep{2016MNRAS.458.3808R,2021MNRAS.501.1677H,2022MNRAS.512.1843H,2023MNRAS.518.4579P}.
The dynamical approach is particularly advantageous in this context due to the inherent faintness of WDs compared to brighter main sequence stars.

Spectroscopic decomposition methods are typically limited to detecting WD-M dwarf binaries, as these systems exhibit a prominent white dwarf signature in their spectra. Notably, the temperature of WDs identified through spectroscopic decomposition tends to be sufficiently high to show a discernible white dwarf component in the spectrum. On the other hand, the UV-excess method faces challenges in identifying sources lacking UV data points, as observed in objects like J035540 and J035916. Additionally, the UV radiation from stellar chromospheres can significantly contaminate the UV flux of candidates \citep{2017ARA&A..55..159L}.
In contrast, the dynamical method proves to be a more robust approach. By utilizing radial velocity curves to derive orbital parameters and employing mass functions, this method allows us to confidently determine the mass of unseen objects.

\section{summary}\label{sec: summmary}

J035540 and J035916 are two binary systems containing compact object candidates identified through meticulous screening of LAMOST LRS, with the visible components being M-type stars. Follow-up spectroscopic observations from P200 augment the radial velocity measurements of two sources.  
We obtain their orbital periods and radial velocity semi-amplitudes from LAMOST LRS and P200 radial velocities, in conjunction with the ZTF light curves. 
The mass functions of the unseen companions are calculated, with $f(M_2)$ being $0.22\pm0.01 M_{\odot}$ for J035540 and $0.16\pm0.01 M_{\odot} $ for J035916.
Mass estimates for the visible M-type stars are derived through SED fitting, which in turn constrains the minimum mass of the invisible companions via mass functions. Given that the invisible components' minimum masses exceed those of the visible stars and that the spectra are consistent with single-lined spectroscopic binaries, we infer that these unseen stars are likely compact objects.

J035540's ZTF light curve is modeled using \texttt{PHOEBE} with a pure ellipsoidal modulation, providing constraints on the system's inclination. Consequently, we obtain a mass of $M_2 = 0.70^{+0.12}_{-0.05} M_{\odot}$ of the compact object. Due to the asymmetric double-peaked nature of J035916's light curve, which exhibits strong model dependence when simulated with a spot model, an analysis based on ellipsoidal modulation is employed. This analysis constrains the light curve amplitude, thereby restricting the orbital inclination and yielding a mass range of $0.57-0.90 M_{\odot}$ for the compact object. 
We measure the $v_{\rm{rot}} \sin i$ of J035916 as $59_{-5}^{+6} \  \mathrm{km\,s^{-1}}$ from the red-side P200 spectrum at the quadrature orbital phase, thereby constraining the orbital inclination as $57_{-13}^{+18}$ degrees. This is consistent with the result from the analytical ellipsoidal model.

The mass estimates suggest that both compact objects are likely WDs. 
These results highlight the efficacy of optical time-domain survey data and dynamical methods in the search for WDs. This approach provides a promising avenue for detecting WDs, even for those extremely faint WDs.
Moreover, this methodology holds the potential for uncovering other compact objects within binaries, including stellar-mass black holes and neutron stars, even during their quiescent states.

\section*{Acknowledgements}

We thank Renbin Yan for helpful discussions. We thank the anonymous referee for constructive suggestions that improved the paper.
This work was supported by the National Key R\&D Program of China under grants 2023YFA1607901 and 2021YFA1600401, the National Natural Science Foundation of China under grants 11925301, 12033006, 12221003, and 12263003, and the fellowship of China National Postdoctoral Program for Innovation Talents under grant BX20230020. 
J.Z.L was supported by the Tianshan Talent Training Program through the grant 2023TSYCCX0101.
We acknowledge the science research grants from the China Manned Space Project.
The observation of P200 (CTAP2023-B0098; PI: Senyu Qi) was supported through the Telescope Access Program (TAP). 

\section*{Data Availability}

This paper uses the data from the LAMOST low-resolution survey. Guoshoujing Telescope (the Large Sky Area Multi-Object Fiber Spectroscopic Telescope LAMOST) is a National Major Scientific Project built by the Chinese Academy of Sciences. Funding for the project has been provided by the National Development and Reform Commission. LAMOST is operated and managed by the National Astronomical Observatories, Chinese Academy of Sciences.
ZTF is a public-private partnership, with equal support from the ZTF Partnership and from the U.S. National Science Foundation through the Mid-Scale Innovations Program (MSIP).



\bibliographystyle{mnras}
\bibliography{ms}








\bsp	
\label{lastpage}
\end{document}